\documentstyle[12pt,aaspp4]{article}

\newcommand\kms{\mbox{ km~s$^{-1}$}}
\newcommand\ppc{\mbox{ pc}}
\newcommand\kpc{\mbox{ kpc}}
\newcommand\Hunits{\mbox{km~s$^{-1}$~Mpc$^{-1}$}}

\newcommand\ep{\epsilon}
\newcommand\th{\theta}
\def\qq#1{q_{3#1}}
\def\refeq#1{eq.~(\ref{eq:#1})}
\def\refeqs#1#2{eqs.~(\ref{eq:#1}) and (\ref{eq:#2})}
\def\reffig#1{Figure~\ref{fig:#1}}
\def\reffigs#1#2{Figures~\ref{fig:#1} and \ref{fig:#2}}
\def\reftab#1{Table~\ref{tab:#1}}

\begin{document}

\title{Gravitational Lensing By Spiral Galaxies}
\author{C. R. Keeton}
\author{C. S. Kochanek}
\affil{Harvard-Smithsonian Center for Astrophysics, MS-51\protect \\
			 60 Garden Street \protect \\
			 Cambridge MA 02138 }
\authoremail{ckeeton@cfa.harvard.edu}
\authoremail{ckochanek@cfa.harvard.edu}

\begin{abstract}

We study gravitational lensing by spiral galaxies, using realistic
models consisting of halo, disk, and bulge components combined to
produce a flat rotation curve.  Proper dynamical normalization of
the models is critical because a disk requires less mass than a
spherical halo to produce the same rotation curve---a face-on
Mestel disk has a lensing cross section only 41\% as large as
a singular isothermal sphere with the same rotation curve.  The
cross section is sensitive to inclination and dominated by edge-on
galaxies, which produce lenses with an unobserved 2-image geometry
and a smaller number of standard 5-image lenses.  Unless the disk
is unreasonably massive, disk+halo models averaged over inclination
predict $\lesssim 10\%$ more lenses than pure halo models.  Finally,
models with an exponential disk and a central bulge are sensitive
to the properties of the bulge.  In particular, an exponential
disk model normalized to our Galaxy cannot produce multiple images
without a bulge, and including a bulge reduces the net flattening
of edge-on galaxies.  The dependence of the lensing properties on
the masses and shapes of the halo, disk, and bulge means that a
sample of spiral galaxy lenses would provide useful constraints
on galactic structure.

\end{abstract}

\keywords{gravitational lensing -- galaxies: spiral --
galaxies: structure}

\section{Introduction}

Simple theoretical models of spherical gravitational lenses predict
that spiral galaxies produce only 10--20\% of gravitational lenses
(Turner, Ostriker \& Gott 1984; Fukugita \& Turner 1991; Maoz \&
Rix 1993; Kochanek 1991, 1993, 1996a).  The prediction is roughly
consistent with observations (for a summary, see Keeton \& Kochanek
1996; Keeton, Kochanek \& Falco 1997b).  Specifically, of $\sim25$
known lenses, only B~0218+357 (O'Dea et al.\ 1992; Patnaik et al.\
1993) is produced by a galaxy unambiguously identified as a distant
spiral galaxy based on its colors and mass-to-light ratio as well
as the presence of HI and molecular gas and strong Faraday rotation
(Patnaik et al.\ 1993; Carilli, Rupen \& Yanny 1993; Browne et al.\
1993; Keeton et al.\ 1997b).  The lens Q2237+0305 (Huchra et al.\
1985), found as part of a redshift survey, is a special case of
lensing by the bulge of a nearby spiral galaxy ($z_l=0.04$).  The
radio ring PKS~1830$-$211 (Rao \& Subrahmanyan 1988; Jauncey et al.\
1991) shows both HI and molecular absorption features (Lovell et
al.\ 1996; Wiklind \& Combes 1996) and thus may have a spiral lens
galaxy, but because the absorption features are at different
redshifts and there is no optical identification of the lens galaxy
this lens is still not understood.  For the remaining lenses, the
lens galaxy is generally more consistent with an early-type galaxy,
with the exception of MG~0414+0534 (Hewitt et al.\ 1992) whose red
color matches no standard galaxy type (Lawrence et al.\ 1995; Keeton
et al.\ 1997b).  

Models of individual lenses and the observed numbers of 4-image
lenses seem to require mean axis ratios somewhat flatter than
expected for early-type galaxies (King et al.\ 1996; Kochanek
1996b; Keeton, Kochanek \& Seljak 1997a).  The apparent
discrepancy may be due to difficulties interpretating the axis
ratios of lens models, which is complicated by the effects of
external tidal shears from neighboring galaxies and clusters
(e.g.\ Hogg \& Blandford 1994; Schechter et al.\ 1997), and by
the possibility that dark halos may be flatter than the light
(e.g.\ Dubinski \& Carlberg 1991).  An alternate possibility is
that spherical models may grossly underestimate the number
of spiral galaxy lenses by not adequately representing real
galaxies, which have not only round halos but also flat disks.
Because errors in estimating the expected number of lenses can
bias inferences about the cosmological model based on the
statistics of gravitational lenses, there is growing interest
in studying lensing by spirals using models that better
represent real galaxies.

There are as yet no treatments of lensing by spiral galaxies
using a model that includes both a realistic disk and a halo.
The spherical models treated spiral galaxies as singular
isothermal spheres (SIS) normalized by their rotation curves,
so they represented diskless, pure halo models.  Models using
ellipsoidal densities (Kassiola \& Kovner 1993; Kormann,
Schneider \& Bartelmann 1994ab; Kochanek 1996b; Keeton et al.\
1997a) can be interpreted as projections of disk galaxies
without halos, although they are not generally viewed as such.
Pure disk models are poor representations of spiral galaxies
because they neglect the dynamically important dark halos that
may not be spherical but are certainly not as flat as disks
(see reviews by Ashman 1992 and Rix 1996).  In addition, pure
disk models with a flat rotation curve predict that the cross
section diverges as the disk becomes edge-on, and that the
divergent cross section is dominated by an image geometry
consisting of two bright images offset from the center of the
galaxy and straddling the projected disk.  Among point image
lenses we see only lenses consisting of two or four images
surrounding the center of the galaxy (see Keeton \& Kochanek
1996 for a summary), and the absence of the ``disk'' image
geometry is direct evidence for rounder halos.  What makes
lensing by realistic spiral galaxies interesting, then, is
not the effects of the thin disk, because the properties of
pure disk models were already understood from studies of
ellipsoidal lenses, but the effects of the halo in
suppressing the divergent cross section and unphysical
image geometry of a pure disk model.

The fact that the observational data are not consistent with
pure disk models means that spiral gravitational lenses will
provide a useful probe of the balance between the disk and
the halo in spiral galaxies.  In our own Galaxy, the constraint
on the local surface mass density of the disk of $(75\pm25)\,
M_\odot\ppc^{-2}$ (Kuijken \& Gilmore 1991; Bahcall, Flynn \&
Gould 1992; also see Sackett 1996b) is one of the weakest
links in understanding the mass distribution of the Galaxy
and interpreting the results of the LMC and Galactic bulge
microlensing searches (e.g.\ Alcock et al.\ 1995).  In external
galaxies, the decomposition of rotation curves between the disk
and the halo is usually degenerate and standard models assume
a ``maximal disk'' to derive lower bounds on the halo
contributions (e.g.\ van Albada \& Sancisi 1986).  Thus any
new constraint on the relative contributions of the disk and
the halo in spiral galaxies has significance well beyond its
particular effects on gravitational lensing.

Recently Maller, Flores \& Primack (1997) and Wang \& Turner
(1997) began to explore the effects of combining a disk with
a halo by embedding a constant surface density, finite radius
disk in a spherical isothermal halo.  Maller et al.\ (1997)
examined the ability of the model to fit B 1600+434 (Jackson
et al.\ 1995), a two-image lens that Jaunsen \& Hjorth (1997)
suggested is a spiral galaxy.  Wang \& Turner (1997) examined
the inclination-averaged cross section to see if the spherical
models systematically underestimate the number of lenses
produced by spirals.  The constant density disk model is
analytically tractable, but the mass density and rotation
curve bear little resemblance to a real galaxy, and the sharp
disk edge introduces peculiar features in the lensing properties.
Here we introduce several simple, physically reasonable models
for lensing by spiral galaxies by combining halo, disk, and
bulge components to produce nearly flat rotation curves.  In
\S2 we describe the halo, disk, and bulge components and
discuss their lensing properties.  In \S3 and \S4 we combine
the components into realistic models and study the effects of
inclination and the shapes and masses of the halo, disk, and
bulge on the lensing cross section, optical depth, and image
geometries.  In \S5 we summarize our results and discuss their
implications for lensing statistics, galactic structure, and
the statistics of damped Ly$\alpha$ absorbers.

\section{Model Components:  Thin Disks and Oblate Halos}

We build realistic models for spiral galaxies by embedding a
thin disk and possibly a central bulge in a dark matter halo.
We can describe both disky and spheroidal components by using
an oblate spheroid with axis ratio $\qq{}$, and then letting
$\qq{} \to 0$ for an infinitely thin disk or $\qq{} \lesssim 1$
for a moderately flattened halo or bulge.  An oblate spheroid
projects to an ellipsoidal density distribution with
projected axis ratio $q=(\qq{}^2 \cos^2 i + \sin^2 i)^{1/2}$,
where $i$ is the standard inclination angle ($i=90^\circ$ is
face-on and $i=0^\circ$ is edge-on).  In the limit of an
infinitely thin disk, a surface mass distribution $\Sigma_3(R^2)
\delta(z)$ projects to an ellipsoid with surface density
\begin{equation}
	\Sigma = {1\over q}\,\Sigma_3 \left(x^2+y^2/q^2\right)
\end{equation}
where $q=|\sin i|$.  Thus the ellipsoidal gravitational lens
models used by Kassiola \& Kovner (1993), Kormann et al.\ (1994ab),
Kochanek (1996b), and Keeton et al.\ (1997a) can be viewed in the
traditional way as models of early-type galaxies (projections of
three-dimensional ellipsoids), or can be reinterpreted as pure
disk models of spiral galaxies (projections of a two-dimensional
disk).  However, the dynamical normalization differs for the two
interpretations; we focus on the disk interpretation and
occasionally discuss its relation to the early-type galaxy
interpretation.

A simple building block for galaxies with flat rotation curves is
the softened, oblate, isothermal density distribution.  The density
and rotation curve for this model are
\begin{eqnarray}
	\rho &=& { v_c^2 \over 4 \pi G \qq{} }\,{ e \over \sin^{-1} e}\,
		{ 1 \over s^2+R^2+z^2/\qq{}^2 }\ , \\
	v_c^2(R) &=& v_c^2 \left\{ 1 - { e \over \sin^{-1} e } { s \over (R^2+e^2 s^2)^{1/2} }
		\tan^{-1} \left[ { (R^2+e^2 s^2)^{1/2} \over \qq{} s } \right] \right\} ,
\end{eqnarray}
where $s$ is a core radius, $e=(1-\qq{}^2)^{1/2}$ is the eccentricity
of the mass distribution, and the model is normalized so that
asymptotically $v_c(R)\to v_c$.  The SIS model corresponds to the
limit $\qq{}=1$ and $s=0$.  The projected surface mass density
in units of the critical surface mass density for lensing is
\begin{equation}
	2\Sigma/\Sigma_{cr} =  b_I \left[ q^2(s^2+x^2)  + y^2 \right]^{-1/2}
	\label{eq:MestSig}
\end{equation}
where $b_I = b_{SIS} e/\sin^{-1}e $, $b_{SIS} = 2\pi(v_c/c)^2 D_{LS}/
D_{OS}$ is the critical radius of a singular isothermal sphere with
rotation velocity $v_c$, and $D_{OS}$ and $D_{LS}$ are comoving
distances from the observer to the source and from the lens to the
source, respectively.  The lensing potential, deflection, and
magnification produced by the lens are
\begin{eqnarray}
	\phi_I(s,\qq{}) &=& x\,\alpha,_x \,+\, y\,\alpha,_y
		\,-\, b_I s\,\ln\left[(\psi+s)^2+(1-q^2)x^2\right]^{1/2}
		+ \mbox{constant} , \\
	\alpha,_x &=& {b_I \over (1-q^2)^{1/2}} \tan^{-1}
		\left[{(1-q^2)^{1/2} x \over \psi+s} \right], \\
	\alpha,_y &=& {b_I \over (1-q^2)^{1/2}} \tanh^{-1}
		\left[{(1-q^2)^{1/2} y \over \psi+q^2 s}\right], \\
	M^{-1} &=& 1 - {b_I \over \psi}
		- {b_I^2 s \over \psi \left[(\psi+s)^2+(1-q^2) x^2\right]}\ ,
\end{eqnarray}
where $\psi^2 = q^2(x^2+s^2) + y^2$.  These equations are identical
(up to an overall normalization factor) to those derived in previous
treatments of the softened isothermal ellipsoid (e.g.\ Kassiola \&
Kovner 1993; Kormann et al.\ 1994a), but the analytic forms are simpler.
The normalization is such that a softened isothermal ellipsoid usually
written as $2\Sigma/\Sigma_{cr} = b_{IE}
\left[s_{IE}^2+r^2(1-\ep\cos2\th)\right]^{-1/2}$ can be written
in the form above by identifying $b_I^2 = b_{IE}^2 (1+q^2)/2$,
$s^2 = s_{IE}^2 (1+q^2)/2q^2$, and $q^2 = (1-\ep)/(1+\ep)$.

In the limit of an infinitely thin disk ($\qq{}\to0$), the isothermal
model becomes a disk with the surface density and rotation curve pair
\begin{equation}
	 \Sigma_M(R,s) = { v_c^2 \over  2\pi G }\,{ 1 \over (R^2 + s^2)^{1/2} }\ , \qquad  
	 v_c^2 (R,s) =  v_c^2 \left[ 1 - { s \over (R^2+s^2)^{1/2}} \right].
\end{equation}
We call this model a softened Mestel disk, because in the limit $s\to0$
it becomes a Mestel (1963) disk, the surface density distribution producing
a flat rotation curve.  The lensing potential of the softened Mestel disk
is $\phi_M(s) = \phi_I(s,\qq{}\equiv 0)$, and the deflection scale
$b_M = 2 b_{SIS}/\pi$ is the limit of $b_I$ as $e\to1$.
Because a disk requires less mass than a spherical distribution to produce
a given rotation velocity, one immediate difference between a Mestel disk
and a singular isothermal sphere is that a face-on Mestel disk has
image separations smaller by $b_M/b_{SIS}=2/\pi=0.64$ and a cross
section smaller by $b_M^2/b_{SIS}^2=4/\pi^2=0.41$.  As the inclination
increases the lensing cross section grows, and it diverges for an
edge-on Mestel disk.  We will study this divergence in \S3.2; the main
result is that the cross section diverges not because
the model neglects the finite thickness of the disk, but rather because
the total mass of the disk diverges.  One way to avoid the divergence
is to smoothly truncate the Mestel disk by using the difference of two
Mestel disk models, $\phi_T(s,a)=\phi_M(s)-\phi_M(a)$ where the truncation
radius $a$ is larger than the core radius $s$.  The surface density of
the truncated Mestel disk model is constant for $R<s$ (rising rotation
curve), declines as $1/R$ for $s<R<a$ (flat rotation curve), and declines
as $1/R^3$ for $R>a$ (Keplerian rotation curve).  The truncated Mestel
disk has a finite mass of $M=(a-s)v_c^2/G$.  Note that if the truncated
model is round rather than flat, its density
$\rho \propto 1/(r^2+s^2)(r^2+a^2)$ is similar to a Jaffe (1983) model,
$\rho \propto 1/r^2(r+a)^2$.

A second useful building block is an unnamed density distribution with
$\rho\sim r^{-4}$ asymptotically.  In terms of the total mass $M$,
the density and rotation curve are
\begin{eqnarray}
	\rho &=& {M s \over \pi^2 \qq{}}\, {1 \over (s^2+R^2+z^2/\qq{}^2)^2}\ , \\
	v_c^2(R) &=& {2 G M \over \pi R} \left\{
		{R^3\over(R^2+e^2 s^2)^{3/2}} \tan^{-1}\left[{(R^2+e^2 s^2)^{1/2}
		\over \qq{} s}\right] - {\qq{} s R^3 \over (R^2+s^2)(R^2+e^2 s^2)}
	\right\} .
\end{eqnarray}
In projection the surface mass density, lensing potential, deflection,
and magnification are
\begin{eqnarray}
	2\Sigma/\Sigma_{cr} &=& b_K^2 q^2 s \left[q^2(s^2+x^2)+y^2\right]^{-3/2} ,
		\label{eq:KuzSig} \\
	\phi_K(s,\qq{}) &=& b_K^2 \ln \left[ (\psi+s)^2 + (1-q^2) x^2 \right]^{1/2}
		+ \mbox{constant}, \\
	\alpha,_x &=& { b_K^2 x \over \psi } { \psi+q^2 s \over (\psi+s)^2+(1-q^2)x^2 }\  , \\
	\alpha,_y &=& { b_K^2 y \over \psi } { \psi+s \over (\psi+s)^2+(1-q^2)x^2 }\  , \\
	M^{-1}    &=& 1 - { b_K^2 q^2 s \over \psi^3 }
		\,-\, { b_K^4 q^2 s \over \psi^3 \left[(\psi+s)^2+(1-q^2)x^2\right] } \nonumber \\
	&& \,-\, { b_K^4 \left[\psi^2(\psi+s)^2-s(2\psi+s)(\psi+q^2 s)^2\right] \over
			\psi^4 \left[(\psi+s)^2+(1-q^2)x^2\right]^2 }\ ,
\end{eqnarray}
where the deflection scale $b_K$ is related to the mass by $M=\pi b_K^2
\Sigma_{cr}$.  In the limit of an infinitely thin disk ($\qq{}\to0$),
the model corresponds to a Kuzmin (1956) or Toomre (1962) Model I disk
and can be used to approximate an exponential disk.  It has the same
mass and central surface density as an exponential disk of the form
$\Sigma(R)=\Sigma_0 e^{-R/s}$, and the rotation curves differ by at
most $16\%$.  A true exponential disk in projection requires numerical
integrals, making it cumbersome to use.

Where a cosmological model is required we adopt $\Omega_0=1$ and $H_0=50$
\Hunits.

\section{Truncated Mestel Disks in Softened Isothermal Halos}

We first consider models consisting of a truncated Mestel disk
embedded in an oblate isothermal halo.  A Mestel disk has a surface
mass density that falls off as $R^{-1}$ while observed spiral galaxies
have luminosity densities that fall off as $e^{-R/R_d}$, so a Mestel
disk cannot represent a galaxy with a constant mass-to-light ratio in
the disk.  Nevertheless, the Mestel disk is interesting to study
because it is the simplest disk system with a flat rotation curve.

\subsection{Normalization of the model}

For simplicity, we let the disk be singular ($s=0$), so its only scale
length is the truncation radius $a_d$.  We place the disk in a softened
isothermal halo and tune the ratio of the halo scale radius $a_h$ to
the disk truncation radius $a_d$ to produce a flat rotation curve;
\reftab{magic} gives typical values of the ratio for a rotation curve
that is flat to better than $2\%$.
The inner rotation curve ($R<a_d$) is supported entirely by the disk, so
this is a ``maximal disk'' model for a spiral galaxy (e.g.\ van Albada \&
Sancisi 1986).  It is not known whether most spiral galaxies have maximal
disks, although it is generally believed that our Galaxy has a disk that
is only $\sim50\%$ of maximal (Bahcall 1984; Kuijken \& Gilmore 1989;
van der Kruit 1989; Kuijken \& Gilmore 1991; Kuijken 1995; but see Sackett
1996b for a recent rebuttal).  We allow for a submaximal disk, i.e.\ for
some of the inner rotation curve to be supported by a dark matter halo,
by embedding the disk+softened halo system in a singular isothermal halo.
The overall lensing model is then
\begin{equation}
\phi = f_d \biggl[ \phi_I(0,\qq{d}) - \phi_I(a_d,\qq{d})
	+ \phi_I(a_h,\qq{h}) \biggr] + (1-f_d) \phi_I(0,\qq{h}),
\end{equation}
where the ``disk fraction'' $f_d$ is the fraction of the inner rotation
curve supplied by the disk, and $\qq{d}$ and $\qq{h}$ are the
three-dimensional axis ratios of the disk and halo, respectively.  An
infinitely thin disk has $\qq{d}=0$ and a spherical halo has $\qq{h}=1$.
The projected axis ratios of the disk and halo are
$q_d=(\qq{d}^2 \cos^2 i + \sin^2 i)^{1/2}$ and
$q_h=(\qq{h}^2 \cos^2 i + \sin^2 i)^{1/2}$.  The model contains the
limits of a pure Mestel disk ($f_d=1$ and $a_d\to\infty$) and a pure
isothermal halo (either $f_d=0$ or $a_d\to0$).  It has no bulge
component and a singular central surface density.

Given a rotation velocity $v_c$, it is convenient to normalize the
length scales by the critical radius $b_{SIS}$ of the SIS model with
the same circular velocity, which yields lensing cross sections in
units of the SIS cross section ($\sigma_{SIS} = \pi b_{SIS}^2$) and
thus indicates whether including the disk increases or decreases the
cross section.  We choose values for the disk and halo axis ratios
$\qq{d}$ and $\qq{h}$, the disk truncation radius $a_d$, and the
disk fraction $f_d$, and finally determine the halo core radius
$a_h$ from the value of $a_h/a_d$ that gives a flat rotation curve
(see \reftab{magic}).

For the physical normalization we can compare the model with the
Galaxy; we use IAU value of the circular velocity $\Theta_0=220 \kms$
(Kerr \& Lynden-Bell 1986) and a consensus value for the solar radius
$R_0=8\kpc$ that is slightly smaller than the IAU value of $8.5\kpc$
(see the review by Reid 1993).  The surface mass density of the disk
at $R_0$ is
\begin{equation}
\Sigma_\odot = 350 f_d {e_d \over \sin^{-1} e_d}
	\left[{\Theta_0 \over 220 \kms}\right]^2
	\left[{8\kpc \over R_0}\right]
	\left[ 1 - { R_0 \over (R_0^2 + a_d^2)^{1/2} } \right]\,
	M_\odot\ppc^{-2},
\end{equation}
where $e_d=(1-\qq{d}^2)^{1/2}$ is the eccentricity of the disk.
Local estimates of the surface mass density of the disk are
$(75\pm25)\, M_\odot\ppc^{-2}$ with more of a consensus toward low
values (Kuijken \& Gilmore 1991; Bahcall et al.\ 1992; also see
Sackett 1996b), so we must choose $a_d/R_0 \gtrsim 0.5$,
and we should reduce the disk fraction $f_d$ if $a_d/R_0 \gtrsim 2$.
The physical scale $a_d/R_0$ is related to the dimensionless ratio
$a_d/b_{SIS}$ appearing in the lens models by
\begin{equation}
	{a_d \over b_{SIS}} =  0.20 {a_d \over R_0} \left[{R_0 \over
		8h_{50}^{-1}\kpc} \right] \left[{220\kms \over \Theta_0}\right]^2
		{ 2 r_H (1+z_l) D_{OS} \over D_{OL} D_{LS} }, \label{eq:abaR}
\end{equation}
where $r_H=c/H_0$ is the Hubble radius, $D_{OL}$, $D_{OS}$, and
$D_{LS}$ are comoving distances to the lens, to the source, and
from the lens to the source, respectively (with $D_{ij}=2 r_H
\left[(1+z_i)^{-1}-(1+z_j)^{-1}\right]$ for $\Omega_0=1$), and
$H_0 = 50 h_{50}\ \Hunits$.  With $f_d=1$, the disk dominates
the inner rotation curve of the model provided $a_d/b_{SIS} \gg 1$,
which is always true because the minimum value of the cosmological
distance ratio is $\sim 10$.

\subsection{The effects of inclination}

\begin{figure}[h]
	\plotone{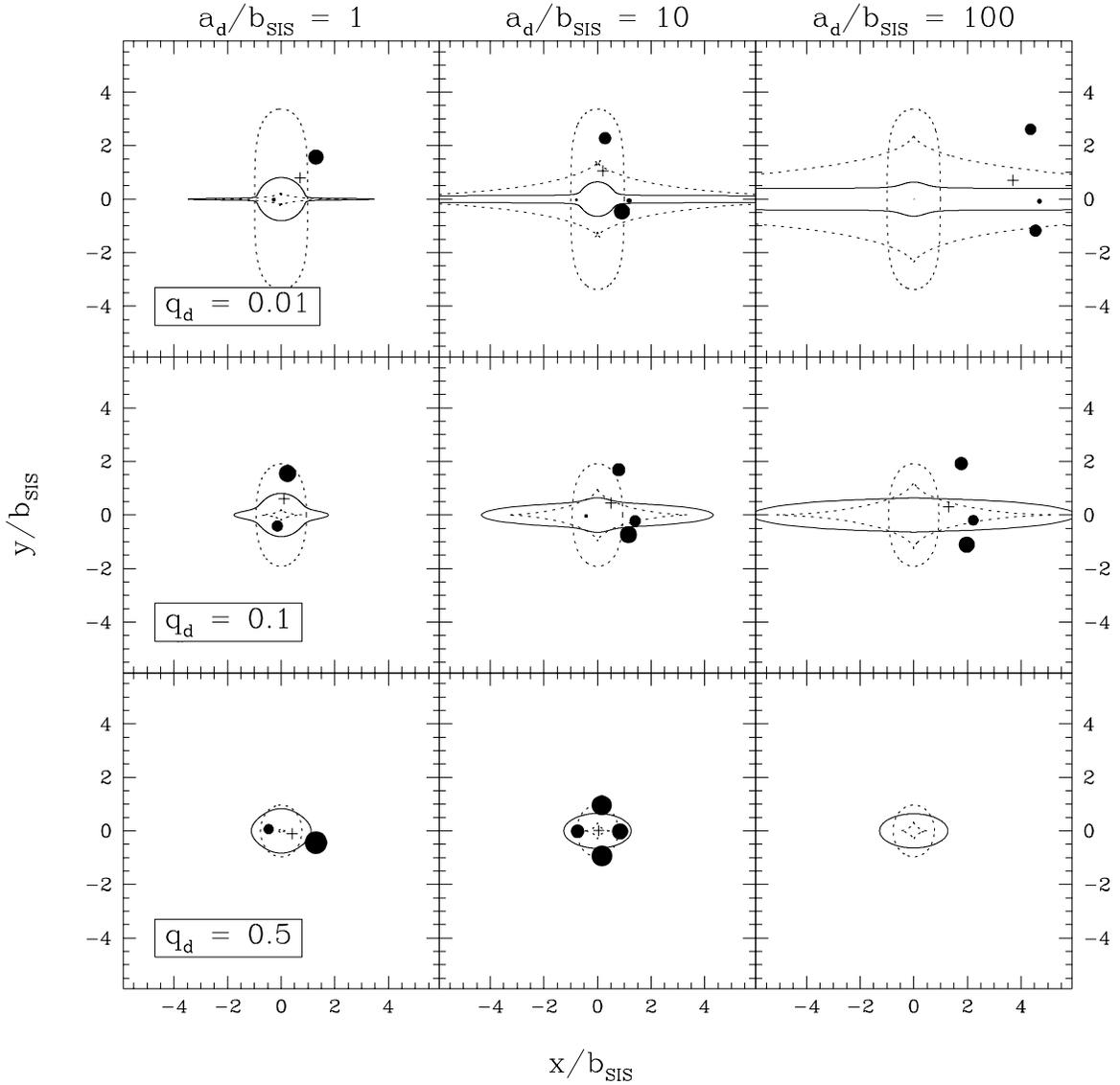}
	\caption{
Sample critical curves, caustics, and image configurations for a
truncated Mestel disk in a spherical isothermal halo.  The projected
disk axis ratio is $q_d=(\qq{d}^2 \cos^2 i + \sin^2 i)^{1/2}$.
In each panel, the solid line is the tangential critical curve in
the image plane and the dotted lines are the tangential and radial
caustics in the source plane.  The three primary image geometries
are illustrated with filled circles indicating images produced by
a source marked with a plus.  The standard 2-image geometry is
shown in the panels with $a_d/b_{SIS}=1$, the 4-image geometry
in the panels with $a_d/b_{SIS}=10$, and the unobserved ``disk''
geometry in the panels with $a_d/b_{SIS}=100$.  The 2-image and
4-image geometries each have an additional image trapped and
demagnified in the singular core of the disk.  The areas of the
circles denote the magnifications.
}\label{fig:mcrit}
\end{figure}

We first consider maximal disk models ($f_d=1$), so the inner rotation
curve is supported entirely by the disk.  The lensing properties of
the model depend strongly on both the inclination (through the axis
ratio $q_d$) and the size (truncation radius $a_d$) of the disk.
\reffigs{mcrit}{mdiv} illustrate the the critical curves, caustics,
image geometries, and cross sections as functions of $a_d$ and $q_d$.

For a face-on galaxy ($q_d=1$), the model is strictly circular and
the only multiple image geometry has three images (with one trapped
and demagnified in the singular core of the disk).  A non-axisymmetric
galaxy would also have a 5-image cross section, but studies of face-on
spiral galaxies indicate that they have axis ratios $b/a \gtrsim 0.7$
(see the review by Rix 1996) so the 5-image cross section would be
small.  We noted in \S2 that a disk requires considerably less mass
than a spherical halo to produce a given rotation velocity, so the
critical radius of a pure Mestel disk ($a_d\to\infty$) is smaller
than the corresponding SIS by $b_M/b_{SIS}=2/\pi$.  Thus a face-on
Mestel disk has a cross section $\sigma_M/\sigma_{SIS}=4/\pi^2=0.41$,
making it less efficient than an SIS at producing multiple images.
As we truncate the Mestel disk, however, the isothermal halo supporting
the outer rotation curve begins to increase the cross section, so
$\sigma_I/\sigma_{SIS}$ depends on the truncation radius $a_d$ and
varies from $4/\pi^2$ for $a_d\to\infty$ to unity for $a_d=0$.

For a modestly inclined galaxy ($q_d\lesssim1$), the tangential critical
line becomes elongated and produces an ``astroid'' caustic corresponding
to standard 4-image geometries (see Schneider, Ehlers \& Falco 1992),
with a fifth image trapped in the singular core of the disk.  As the
inclination increases ($q_d$ decreases), the tangential critical line
becomes even more elongated and the astroid caustic pierces the radial
caustic.  The region inside the astroid caustic but outside the radial
caustic corresponds to a configuration of three images on one side of
the center of the galaxy.  The middle image fades as the inclination
increases ($q_d$ decreases), resulting in a geometry with two bright
images off to one side of the galactic center and straddling the
projected disk.  This image geometry, which we refer to as the ``disk''
image geometry, has not been observed.

For a nearly edge-on galaxy ($q_d \ll 1$) the tangential critical line
consists of a central round region with a narrow ``spike'' extending
out the $x$-axis, and the cross section diverges.  There are two
elements of the divergence.  The first is the divergence of the
radial caustic as $q_d\to0$ and the surface mass density becomes a
line density.  For a disk with a core radius $s$ that is small
compared with the disk truncation radius $a_d$ and the halo scale
radius $a_h$, the radial caustic is determined entirely by the
central part of the disk and is independent of $a_d$ and $a_h$.
The radial caustic moves up the $y$-axis as $|\ln q_d|$ and the
3-image cross section diverges logarithmically.  The divergence is
unobservable because most of the large cross section corresponds to
image geometries where the fluxes differ by orders of magnitude.
This is analogous to the divergent cross section of a point mass
lens (see Schneider et al.\ 1992), which is formally infinite only
because it allows images to pass arbitrarily close to a singular
mass distribution and to be arbitrarily faint.  In practice spiral
galaxy disks are observed to have a finite thickness; for example
Guthrie (1992) found a mean axis ratio of $\qq{d}=0.11$ in a sample
of edge-on spiral galaxies.  A finite disk thickness prevents the
mass density from becoming a singular line density and hence the
cross section from diverging.

\begin{figure}[h]
	\plotone{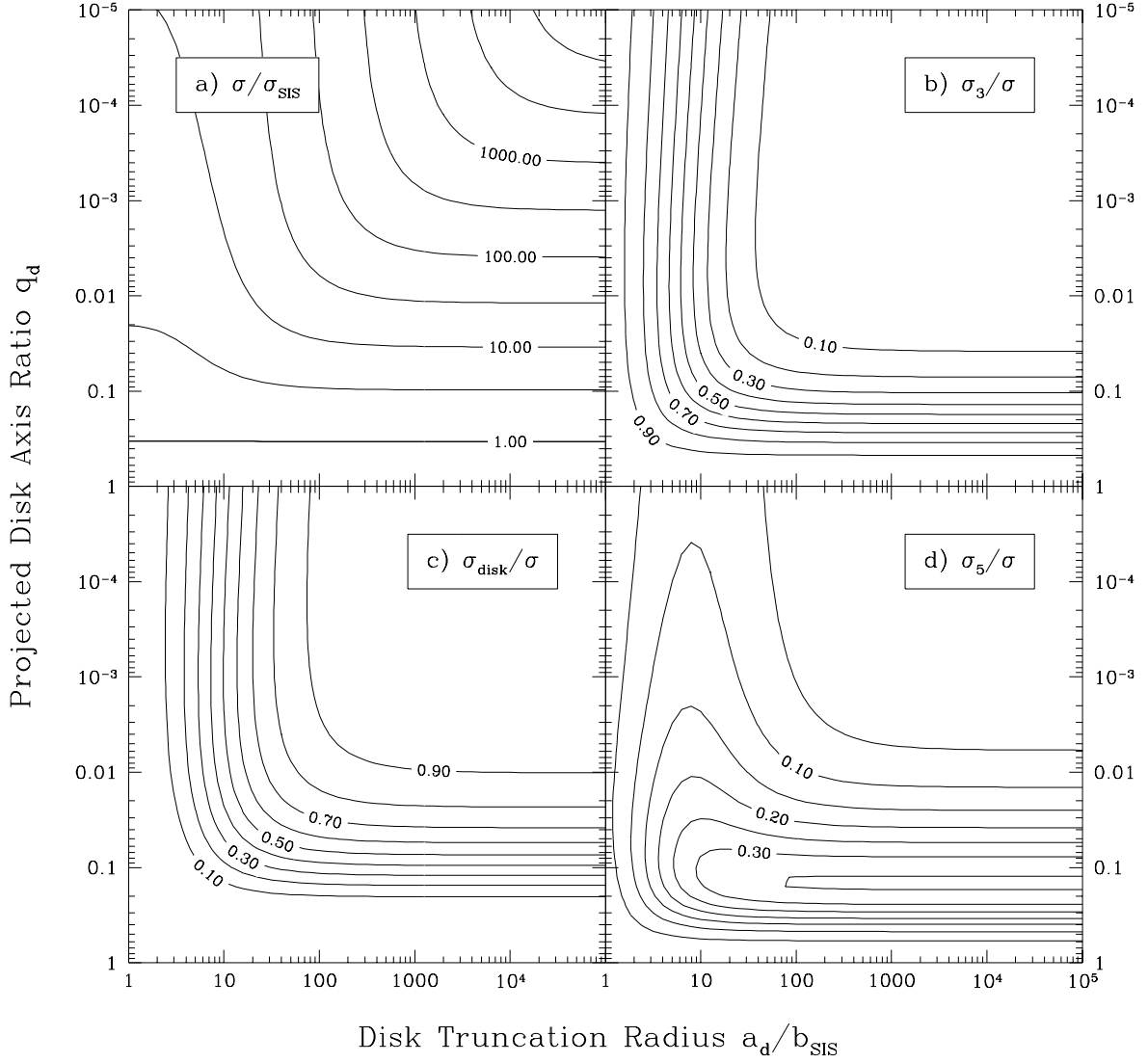}
	\caption{
Cross sections for a truncated Mestel disk in a spherical isothermal
halo, as a function of the disk truncation radius $a_d$ and the
projected disk axis ratio $q_d = (\qq{d}^2 \cos^2 i + \sin^2 i)^{1/2}$.
These are the cross sections for the critical curves and caustics
depicted in Figure 1.
(a)  The total cross section, with contours spaced logarithmically.
(b)--(d)  The branching ratios, or fractions of the total cross section,
corresponding to 3-image geometries, ``disk'' image geometries, and
5-image geometries, respectively.  In (b)--(c) the contour spacing
is 0.1, and in (d) the contour spacing is 0.05.
}\label{fig:mdiv}
\end{figure}

The second element is the divergence of the astroid caustic as
the mass of the disk diverges.  In the limits of an edge-on galaxy
($q_d=0$) or a pure Mestel disk ($a_d \to\infty$) the asymptotic
cross sections are
\begin{equation}
	{\sigma_{astr}\over\sigma_{SIS}} \simeq \cases{
		(4/\pi^2)\, q_d^{-1} & $a_d/b_{SIS} \gg q_d^{-1} \gg 1$ , \cr
		(4/\pi)\, a_d/b_{SIS} & $q_d^{-1} \gg a_d/b_{SIS} \gg 1$ ,
	}
\end{equation}
where $\sigma_{astr}=\sigma_{disk}+\sigma_5$ is the area of the
astroid caustic.  Results for intermediate regimes are shown in
\reffig{mdiv}.  An edge-on disk ($q_d=0$) has an astroid cross
section $\sigma_{astr} \propto a_d \propto M_d$, where $M_d$ is
the disk mass, so that the astroid cross section is finite
provided $M_d$ is finite.  The astroid cross section of a pure
Mestel disk diverges because the mass diverges, not because the
disk is infinitely thin.  Although the cross section is finite
for a truncated disk, it can still be quite large compared to
an SIS model.  Because most of the astroid lies outside the
radial caustic, a nearly edge-on disk is dominated by the ``disk''
image geometry.

In order to produce a realistic disk model and to avoid the
unphysical logarithmic divergence of the radial caustic, we
henceforth give the disk a finite thickness by making it an
oblate spheroid with a small but non-zero $\qq{d}$.  Spheroids
are not ideal representations of the exponential vertical
structure of disks, but because the details of the disk
thickness matter only for inclinations with $\sin i \lesssim
\qq{d}$ we use spheroids for analytic simplicity.

\subsection{The effects of disk and halo masses and shapes}

We can characterize the expected contribution of spiral galaxies to
lensing statistics by computing cross sections and optical depths
averaged over inclination.  In doing so we neglect the magnification
bias that, if included, would tend to reduce the inclination dependence
of the 5-image and disk geometry cross sections because the mean
magnification is higher when the cross section is lower.  Total
probabilities, however, stay roughly proportional to the optical
depth (see Wallington \& Narayan 1993; Kochanek 1996b; Keeton et
al.\ 1997a).

\begin{figure}[h]
	\plotone{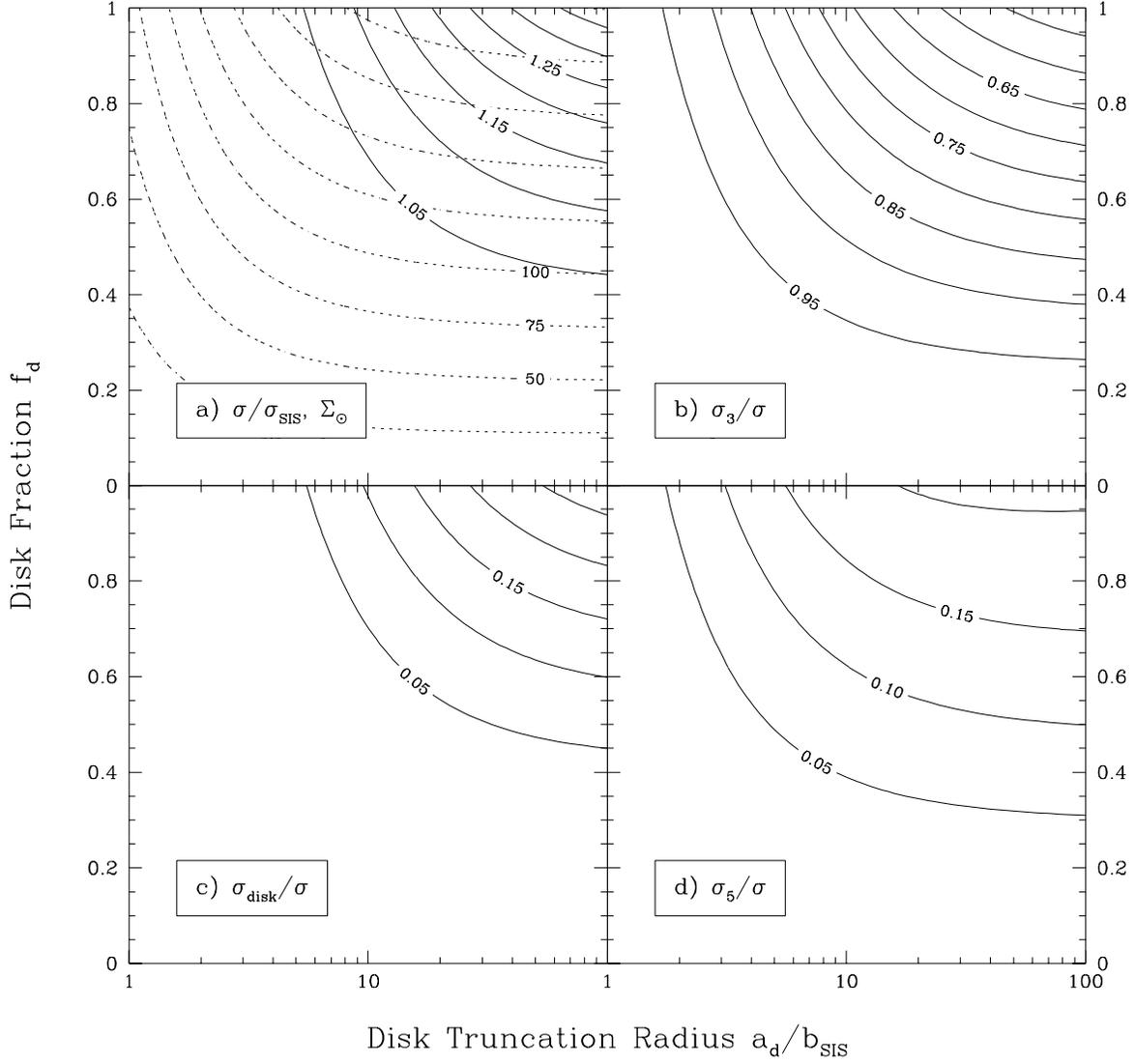}
	\caption{
The inclination-averaged cross section and branching ratios for a
truncated Mestel disk with a finite thickness $\qq{d}=0.03$ in a
spherical halo ($\qq{h}=1$), as a function of the disk truncation
radius $a_d$ and the disk fraction $f_d$.  The contour spacing is
0.05.  Panel (a) also shows dotted lines indicating contours of the
disk surface mass density at $R_0=8\kpc$ for a circular velocity
$\Theta_0=220\kms$, for a source at $z_s=2$ and a lens galaxy at
$z_l=0.5$.  The contour spacing is $25\,M_\odot \ppc^{-2}$.  The
local estimate for the Galaxy is $\Sigma_\odot=(75\pm25)\,
M_\odot\ppc^{-2}$.
}\label{fig:msig}
\end{figure}

\begin{figure}[h]
	\plotone{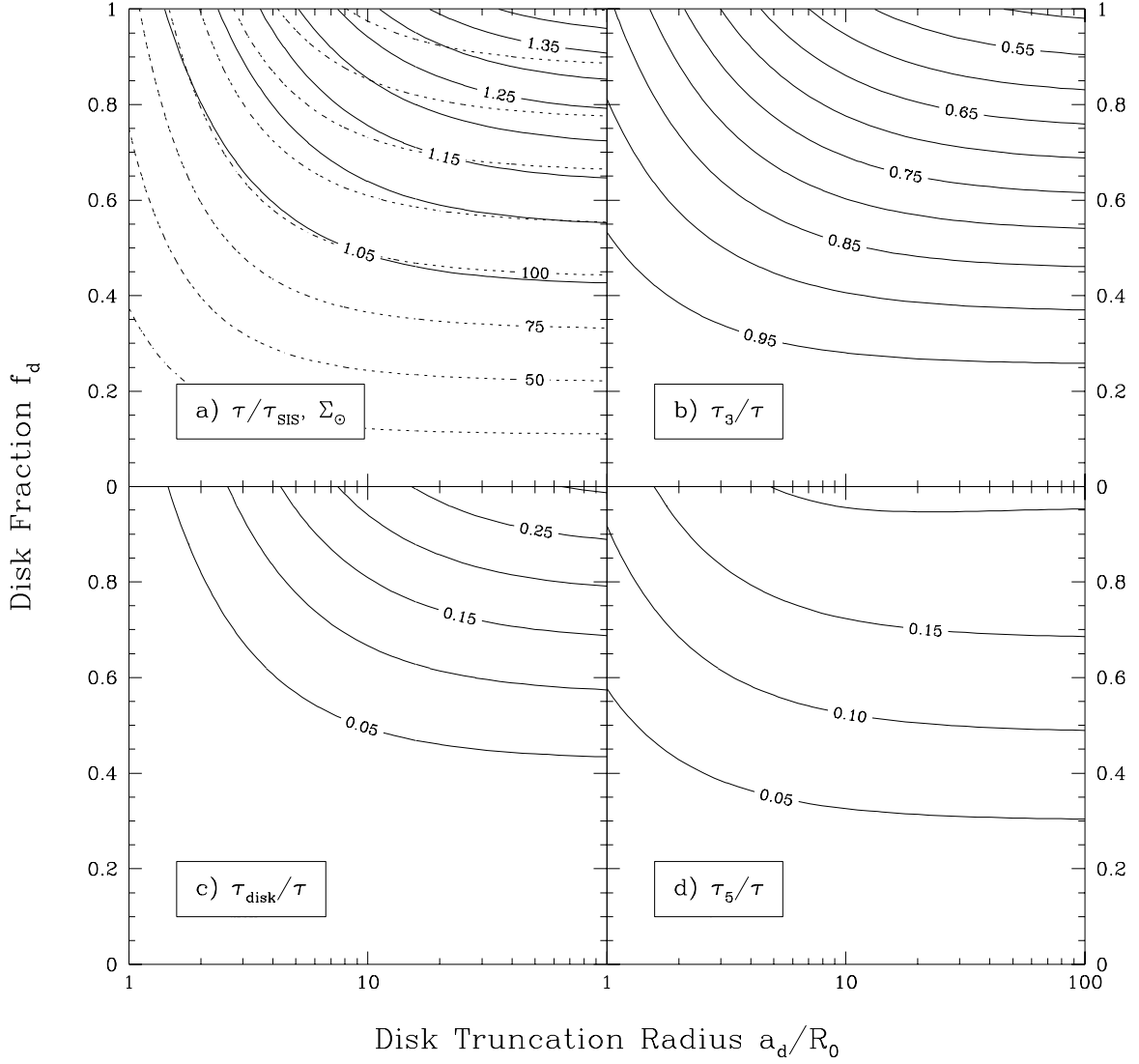}
	\caption{
The inclination-averaged optical depth and branching ratios for the
model in Figure 3, where the optical depth is computed by integrating
over lens redshift for a source at redshift $z_s=2$.  The contour
spacing is 0.05.  The solar radius is $R_0=8\kpc$.
}\label{fig:mtau}
\end{figure}

\reffig{msig} shows the inclination-averaged cross section as a
function of the disk truncation radius $a_d/b_{SIS}$ and the disk
fraction $f_d$, for a disk with thickness $\qq{d}=0.03$ in a
spherical halo.  \reffig{mtau} shows the corresponding optical
depth as a function of $a_d/R_0$.  Note that $a_d/R_0$ is related
to $a_d/b_{SIS}$ by the redshift-dependent factor given in
\refeq{abaR}, so integrating over redshift to obtain the optical
depth is equivalent to integrating over $a_d/b_{SIS}$.  Somewhat
surprisingly, although the face-on cross section is small and
the edge-on cross section is large, the inclination-averaged cross
section and optical depth (in units of the SIS values) are near
unity.  In other words, the disk+halo model does not significantly
increase the number of lenses expected from spiral galaxies over
the simple spherical SIS model.  The number of lenses can be
increased by $\sim40\%$ only if $a_d$ is large and $f_d$ is near
unity, corresponding to a disk that is much more massive than in
the Galaxy (Figures \ref{fig:msig}a and \ref{fig:mtau}a).  Moreover,
many of the additional lenses have the unobserved ``disk'' image
geometry (Figures \ref{fig:msig}c and \ref{fig:mtau}c).  Thus the
disk+halo models that predict significantly more lenses than the
spherical models are physically implausible, while models with a
reasonable disk mass increase the total number of expected lenses
by $\lesssim 10\%$.

\begin{figure}[h]
	\plotone{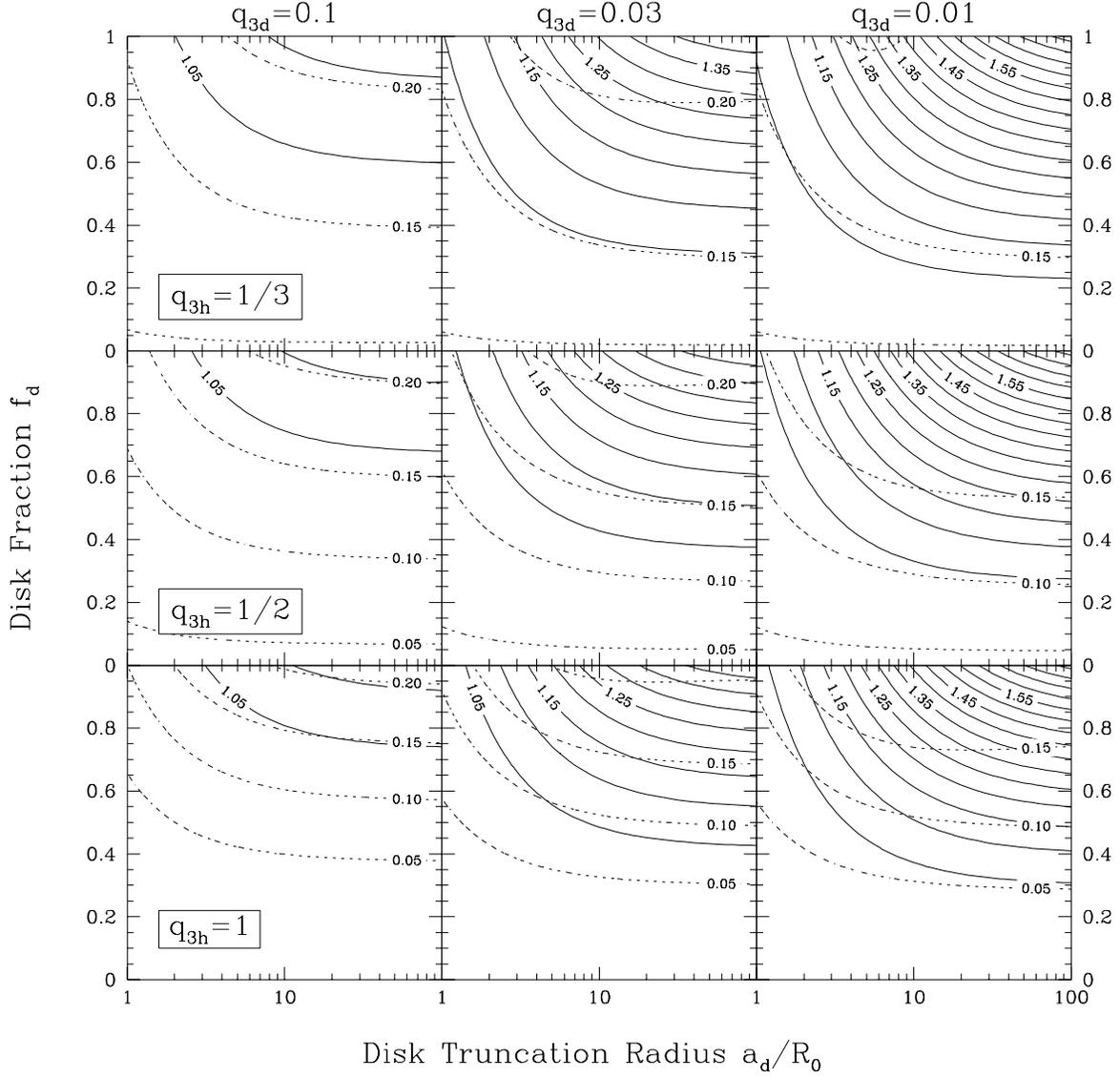}
	\caption{
Contours of the inclination-averaged optical depth $\tau/\tau_{SIS}$
(solid) and 5-image lens fraction $\tau_5/\tau$ (dotted) for a Mestel
disk in an isothermal halo, for various values of the disk thickness
$\qq{d}$ and halo oblateness $\qq{h}$.  The contour spacing is 0.05.
Moving up or to the right in the diagram increases the effective
flattening of the galaxy.
}\label{fig:mtauq3}
\end{figure}

Evidence from observations and from N-body simulations suggests
that the dark halos of spirals are not spherical (see the reviews
by Rix 1996 and Sackett 1996a), so in \reffig{mtauq3} we consider
the effects of of flattening the halo.  We also consider making
the disk both thicker and thinner.  Changing the disk thickness
has little effect on the 5-image lens fraction ($\tau_5/\tau$)
but significantly changes the ``disk'' lens fraction ($\tau_{disk}/
\tau$, not shown) and the total optical depth.  This is because
a thicker disk (larger $\qq{d}$) rules out the thin edge-on models
that increase the cross section with numerous ``disk'' lenses.
By contrast, making the halo oblate has little effect on the total
optical depth but significantly changes the 5-image and ``disk''
lens fractions.  This makes sense because flattening the halo
increases the net flattening of the system, thus causing more
5-image and ``disk'' lenses, while reducing the halo mass needed
to produce the same rotation curve.  Apparently the two effects
conspire to keep the total optical depth unchanged, suggesting
that flattening the halo---even as much as 3:1---does little to
increase the total number of lenses.  With any reasonably shaped
halo the only way to increase the number of lenses by $\sim50\%$
compared to the simple SIS model is to let the mass of the system
be dominated by the disk.

\begin{figure}[h]
	\plotone{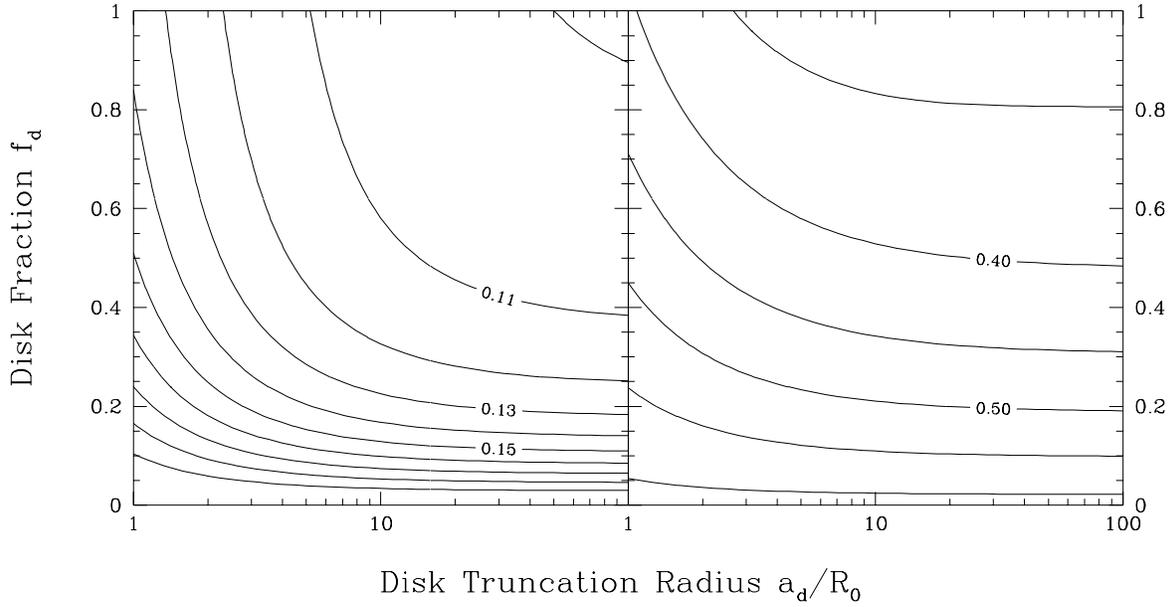}
	\caption{
Properties of the distribution of 5-image lenses for a truncated
Mestel disk with thickness $\qq{d}=0.03$ in a 2:1 flattened isothermal
halo.
Left:  The median value of $\sin i$ for the distribution of 5-image
lenses with inclination [$d\tau_5/d(\sin i)$], with contour spacing
0.01.  Half of all 5-image lens galaxies should be at least this
close to edge-on.
Right:  The axis ratio $q_{SIE}$ of the singular isothermal ellipsoid
producing the same fraction of 5-image lenses ($\tau_5/\tau$) as
the spiral galaxy model, with contour spacing 0.05.
}\label{fig:mqchar}
\end{figure}

In addition to studying the expected number of lenses, we can also
study their distribution with inclination.  We use the optical depth
distribution $d\tau_5/d(\sin i)$ to estimate the number of 5-image
lenses produced by a galaxy with inclination $i$.  Although the
distribution of spiral galaxies should be uniform in $\sin i$, we
know from \S3.2 that the optical depth is dominated by nearly edge-on
systems.  \reffig{mqchar}a shows the median value of $\sin i$ for a
model with $\qq{d}=0.03$ and a 2:1 flattened halo.  The results
depend on $\qq{d}$ and $\qq{h}$, but in most of the parameter
space we have considered the median value is less than
$\sin 10^\circ = 0.17$.  In other words, because of the strong
dependence of the cross section on the inclination, more than
half of 5-image lenses produced by spiral galaxies should come
from systems with $|i|<10^\circ$, i.e.\ systems within $10^\circ$
of edge-on.  The cross section for 5-image lenses is strongly
correlated with the cross section for ``disk'' lenses because
both image geometries are associated with the astroid caustic.
The edge-on galaxies that produce most of the 5-image lenses
also produce ``disk'' lenses, although the ratio of ``disk''
to 5-image lenses depends on the disk thickness.  For example,
a disk with thickness $\qq{d}=0.03$ and a reasonable disk mass
produces about half as many ``disk'' lenses as 5-image lenses
(see \reffig{mtau}c--d), while a thicker disk eliminates thin
edge-on models and hence reduces the number of ``disk'' lenses.

We noted in \S2 that disk lens models are closely related to ellipsoid
lens models for early-type galaxies.  One way to think about the
relation is to compare the fractions of 5-image lenses they produce.
\reffig{mqchar}b shows the axis ratio of the singular isothermal
ellipsoid (SIE) that produces the same fraction of 5-image lenses
(i.e.\ the same $\tau_5/\tau$) as the inclination-averaged disk+halo
model with $\qq{d}=0.03$ and $\qq{h}=0.5$.  The results depend on
$\qq{d}$ and $\qq{h}$, but most models with a plausible disk mass
have $q_{SIE}$ between $\sim0.4$ and $\sim0.6$, with the flatter halos
giving the lower values.  In other words, in terms of the 5-image
lens fraction, inclination-averaged spiral galaxies correspond
roughly to E4--E6 elliptical galaxies.  One key difference,
though, is that the elliptical galaxies would not produce lenses
with the ``disk'' image geometry.

\section{Kuzmin Disks in Softened Isothermal Halos}

Real spiral galaxies have exponential disks and central bulges,
so the inner regions are not well described by the Mestel disk
models of \S3.  We now use a Kuzmin disk as an approximation to
an exponential disk, and we embed the disk in an isothermal halo
to obtain the lensing model
\begin{equation}
	\phi(\mbox{disk+halo}) = \phi_K(R_d,\qq{d}) + \phi_I(a_h,\qq{h}) ,
\end{equation}
where $R_d$ is the scale length of the exponential disk, and $a_h$
is the scale radius of the halo.  In \S3 we examined the effects
of varying the disk thickness $\qq{d}$ and the halo oblateness
$\qq{h}$, so for simplicity we use a thin disk with $\qq{d}=0.03$
and a 2:1 flattened halo ($\qq{h}=0.5$).  We have four remaining
parameters (the scale lengths $R_d$ and $a_h$, the disk mass $M_d$,
and the asymptotic circular velocity $v_c$), but by requiring that
the disk+halo rotation curve be as flat as possible we can fix two
ratios,
\begin{equation}
	G M_d / R_d v_c^2 = 2.577 \qquad\mbox{and}\qquad
	a_h/R_d = 2.229 . \label{eq:Kuznorm}
\end{equation}
With these constraints, the rotation curve starts at zero, rises
to a peak $6\%$ above $v_c$ at $R=1.8 R_d$, falls to a minimum
$0.6\%$ below $v_c$ at $R=12.7 R_d$, and then slowly asymptotes
to $v_c$.  We can then normalize the model using the Galaxy as
in \S3, which we take to have scale length $R_d=3.5\kpc$ (see
Sackett 1996b).  The local surface density and the total disk
mass are then
\begin{eqnarray}
\Sigma_{\odot} &=& 85
	\left[{\Theta_0\over 220\kms}\right]^2
	\left[{3.5\kpc\over R_d}\right]
	\left[{1+(8/3.5)^2 \over 1+(R_0/R_d)^2}\right]^{3/2}
	M_\odot\ppc^{-2} , \\
M_d &=& 10^{11}
	\left[{\Theta_0\over 220\kms}\right]^2
	\left[{R_d\over 3.5\kpc}\right] M_\odot ,
\end{eqnarray}
so the disk is significantly more massive than the estimate of
$6\times10^{10}\,M_\odot$ for our Galaxy (e.g.\ Bahcall 1986;
Binney \& Tremaine 1987), but the local surface mass density
is consistent with the estimates of $(75\pm25)\, M_\odot
\ppc^{-2}$ (Kuijken \& Gilmore 1991; Bahcall et al.\ 1992).

Combining the surface mass densities for the halo and disk from
\refeqs{MestSig}{KuzSig}, the central surface mass density in units
of the critical density for lensing is
\begin{equation}
	\kappa_0 = 
		{1\over2}\left[{b_h \over q_h a_h}+{b_d^2\over q_d R_d^2}\right] .
		\label{eq:kap0}
\end{equation}
A circular system is ``supercritical,'' i.e.\ can produce multiple
images, only if $\kappa_0>1$ (see Schneider et al.\ 1992).  If we
normalize the disk and halo as above and consider source and lens
redshifts $z_s=2$ and $z_l=0.5$ then we have
\begin{equation}
	a_h/b_{SIS} = 2.784 , \qquad
	R_d/b_{SIS} = 1.249 , \qquad \mbox{and} \qquad
	b_d/b_{SIS} = 1.431 ,
	\label{eq:ratios}
\end{equation}
so $\kappa_0$ does not exceed unity until $|\sin i| < 0.83$ or
$|i| < 56^\circ$.  With a true exponential disk the same analysis
yields $a_h/b_{SIS}=2.747$ and $b_d/b_{SIS}=1.338$, so $\kappa_0$
does not exceed unity until $|i| < 48^\circ$.  Thus nearly face-on
systems (normalized to our Galaxy) are subcritical and cannot
produce multiple images.  Modestly inclined systems are just
barely supercritical, so although they can produce multiple images
their cross section for lensing is small.  

Thus the low central surface density of the disk means that the
bulge plays a crucial role in gravitational lensing by spiral
galaxies.  The bulge of our galaxy is well described by a de
Vaucouleurs (1948) $r^{1/4}$ law (e.g.\ Bahcall 1986), but
lensing by a de Vaucouleurs model is impractical because it
requires five independent numerical integrals at every position.
We could approximate the bulge with a modified Hubble profile
$\rho \propto [1+(r/a)^2]^{-3/2}$, but the bulge mass $M_b(r)$
would diverge logarithmically and we would be unable to
characterize the bulge by its mass.  So as a simple way to
examine the qualitative effects of a central bulge with a
finite mass, we use a bulge with the $\rho\sim r^{-4}$ profile
discussed in \S2.  The total lens model is then
\begin{equation}
	\phi = \phi_K(R_d,\qq{d}) + \phi_I(a_h,\qq{h}) + \phi_K(a_b,\qq{b}) ,
\end{equation}
where $a_b$ is the scale radius and $\qq{b}$ the axis ratio for
the bulge.  For simplicity we assume a fixed value for $\qq{b}$.
A self-consistent disk+bulge model requires a flattened bulge
(e.g.\ Monet, Richstone \& Schechter 1981), so without attempting
to build a self-consistent model we fix $\qq{b}=0.5$ for a 2:1
flattened bulge.

\begin{figure}[h]
	\plotone{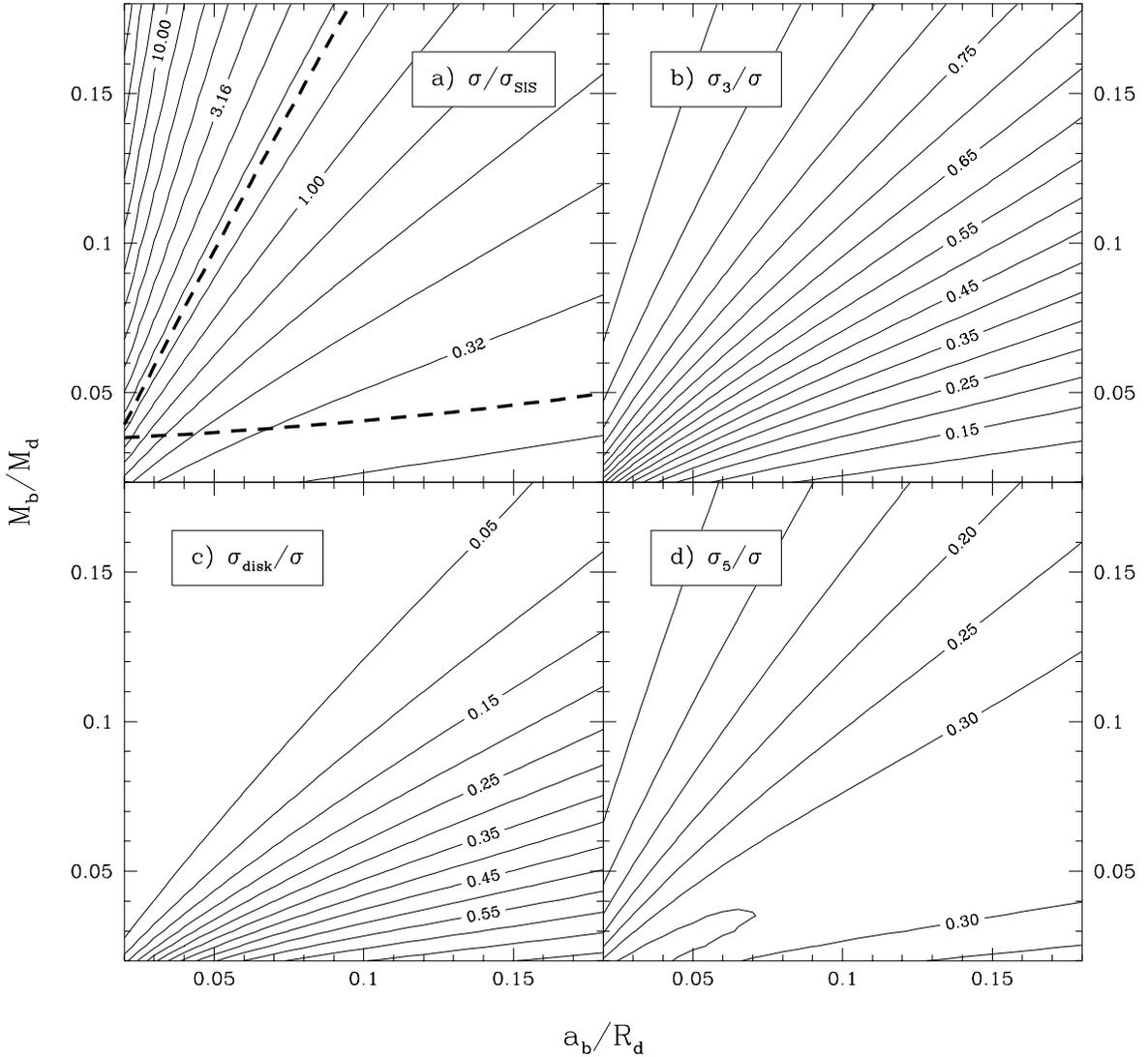}
	\caption{
The inclination-averaged cross section and branching ratios for
a Kuzmin disk in an isothermal halo, as a function of the bulge
to disk scale length ratio $a_b/R_d$ and mass ratio $M_b/M_d$,
where the disk and halo properties are held fixed.  The halo
and bulge are both 2:1 flattened.  (a) The contours are spaced
logarithmically.  The heavy dashed lines indicate the range of
parameters that give a reasonable rotation curve.  Above the
upper line, the bulge causes a central peak in the circular
velocity that is at least 20\% higher than the asymptotic
velocity $v_c$.  Below the lower line the bulge cannot support
the inner rotation curve and the velocity at half a disk scale
length is at least 20\% lower than the asymptotic velocity, i.e.\
$v_c(R_d/2) < 0.8 v_c$.  (b)--(d) The contour spacing is 0.05.
}\label{fig:kuzsig}
\end{figure}

\reffig{kuzsig} shows the inclination-averaged cross section
for lensing as a function of the bulge mass $M_b$ and scale
length $a_b$.  The bulge strongly affects the cross section,
primarily by controlling the central surface density.  With
a diffuse, low mass bulge (large $a_b$ and small $M_b$), the
system is barely supercritical and the cross section is nonzero
but small.  As the bulge becomes massive and concentrated ($a_b$
decreases and $M_b$ increases), the cross section increases
dramatically.  The divergent cross section is misleading, though,
because it is a point mass divergence (see Schneider et al.\
1992).  Including magnification bias and limits on detectable
flux ratios would reduce the cross section to a reasonable value.
In addition to increasing the central surface density, the bulge
also circularizes the center of the galaxy.  As a result, the
5-image and ``disk'' lenses that are associated with a flattened
system become less significant as the bulge becomes more dominant
(see \reffig{kuzsig}c--d).  Thus the bulge can regulate the
numbers of 5-image and ``disk'' lenses, and an analysis of the
distribution of image geometries in a sample of spiral lenses
must account for this effect.

Not all of the bulge parameter space in \reffig{kuzsig} is
physically reasonable.  A concentrated bulge produces an
unphysical mass distribution whose rotation curve has a strong
central peak; in \reffig{kuzsig}a we show where the peak in the
rotation curve due to the bulge is 20\% higher than the asymptotic
circular velocity $v_c$.  Conversely, a low mass bulge cannot
support the inner rotation curve; in \reffig{kuzsig}a we also
show where the circular velocity at $R=R_d/2$ is only 80\% of
the asymptotic value.  If we require that the inner rotation
curve not deviate by more than 20\% from $v_c$, then
\reffig{kuzsig} shows that the inclination-averaged cross
section remains comparable to or smaller than the SIS result.

\section{Discussion}

The traditional approach to gravitational lensing by spiral
galaxies (Turner et al.\ 1984; Fukugita \& Turner 1991; Maoz
\& Rix 1993; Kochanek 1991, 1993, 1996a) neglected the disk
and used the dark halo alone to estimate that only 10--20\%
of gravitational lenses should be produced by spiral galaxies.
Recent ellipsoidal lens models (Kassiola \& Kovner 1993;
Kormann et al.\ 1994ab; Kochanek 1996b; Keeton et al.\ 1997a)
can be reinterpreted as projections of disks to show that
pure disk models viewed nearly edge-on can sharply increase
the number of lenses compared with the pure halo models, but
that most of the additional lenses have an unphysical ``disk''
image geometry with two bright images off to one side of the
galactic center and straddling the projected disk.  Thus it
is important to have a halo to regulate the unphysical effects
of a disk.  Maller et al.\ (1997) and Wang \& Turner (1997)
recently considered a constant surface density, finite radius
disk in a spherical isothermal halo, but their model had a
mass distribution and rotation curve very different from real
galaxies.  We have constructed physically plausible models by
combining disk, halo, and bulge components normalized to produce
a nearly flat rotation curve.  We considered two classes of
models:  a truncated Mestel (1963) disk, which has dark matter
in the disk, in an isothermal halo; and a Kuzmin (1956) disk
as an approximation to an exponential disk, with a central
bulge and an isothermal halo.  These models reveal four
distinctive features of lensing by spirals.

\begin{enumerate}

\item Proper dynamical normalization of the models is important.
A disk requires less mass than a spherical halo to produce the
same rotation curve, so a disk model can have a lensing cross
section significantly smaller than the corresponding halo model.
For example, the cross section of a face-on Mestel disk is only
41\% of the SIS cross section, and the cross section of a
face-on exponential disk in an isothermal halo can be small or
even zero (depending on the mass of the bulge).

\item The disk makes the lensing effects sensitive to the
inclination.  The cross section increases dramatically with
inclination and is dominated by nearly edge-on models.  For
example, more than half of all 5-image lenses produced by a
Mestel disk galaxy come from galaxies within $10^\circ$ of
edge-on.  The cross section for 5-image lenses is correlated
with the cross section for lenses with the unobserved ``disk''
image geometry, so edge-on galaxies also produce significant
numbers of ``disk'' lenses, although the disk thickness and
a bulge offer ways to control the ratio of the two geometries.

\item Despite the inclination effects, disk+halo models
averaged over inclination do not significantly increase the
cross section compared with pure halo models.  The constant
density disk model of Wang \& Turner (1997) predicted
qualitatively that the disk can increase the cross section
by at most $\sim50\%$, and our models normalized to produce
a given rotation curve restrict the increase to $\lesssim
10\%$.  Our models show that increasing the cross section
or optical depth by even $\sim50\%$ requires a very massive
disk that dominates the dark halo, in conflict with
observations that dark halos contribute significantly to
spiral galaxy dynamics (see the review by Ashman 1992).
Our conclusion is insensitive to the shape of the halo,
even for a halo that is 3:1 flattened.  Flattening of the
halo changes the fraction of lenses with 5-image or
``disk'' image geometries but has little effect on the
total number of lenses.

\item A central bulge plays a crucial role in lensing by
spiral galaxies with exponential disks, because an exponential
disk normalized to our Galaxy has a face-on central surface
density too small to produce multiple images.  The bulge
raises the face-on central surface density enough to allow
multiple imaging, with a diffuse bulge producing a small
lensing cross section and a concentrated bulge producing
a large cross section.  The bulge also circularizes the center
of the galaxy, diluting the effects of an edge-on disk and
reducing the number of 5-image and ``disk'' lenses.  Replacing
the bulge with a triaxial bar would give a face-on galaxy a
small 5-image cross section but would otherwise have little
effect.

\end{enumerate}

Our calculations neglected magnification bias so that we could
perform large parameter surveys.  Because mean magnifications
tend to be large when cross sections are small, magnification
bias would tend to reduce the inclination dependence of the
5-image and ``disk'' image cross sections (see Schneider et
al.\ 1992).  In addition, magnification bias would increase
the ratio of 5-image and ``disk'' lenses (which have larger
mean magnifications) to 3-image lenses (which have smaller
mean magnifications).  Thus magnification bias is important
for comparison to any observational sample.  It should not,
however, significantly affect the total cross section or the
ratio of ``disk'' lenses to 5-image lenses.

These results suggest that lensing by spiral galaxies can
provide a new constraint on the structure of spiral galaxies.
At present the balance between disk and halo masses and the
shapes of halos are poorly known.  For example, the
contribution of the Galactic disk to the rotation curve is
not precisely known (Kuijken \& Gilmore 1991; Bahcall et al.\
1992; also see Sackett 1996b), and it has been suggested that
explaining the microlensing optical depth toward the Galactic
bulge requires a disk that is heavier and closer to maximal
than conventionally thought (e.g.\ Alcock et al.\ 1995).  
A sample of spiral galaxy lenses would constrain the relative
masses and shapes of disks and halos, particularly if combined
with HST images to determine the inclination of the disk.
Discovering a lens with the ``disk'' image geometry would
strongly constrain the disk surface mass density, while the
continued absence of ``disk'' lenses would rule out disks
with a surface density significantly higher than our Galaxy.
Unfortunately, a sample of spiral lenses may be difficult
to obtain because they should contribute only 10--20\% of
all lenses, and because the small image separations and
extinction in the lens galaxies may bias optical surveys
against finding them.

Our results also have implications for the predicted
correlation between gravitational lensing and damped
Ly$\alpha$ absorption.  Damped Ly$\alpha$ systems are
thought to be associated with galactic disks (Wolfe 1988,
1995) and may thus produce lensing effects in background
quasars.  Bartelmann \& Loeb (1996) and Smette, Claeskens
\& Surdej (1997) have pointed out that lensing can affect
the statistics of damped Ly$\alpha$ absorbers through
magnification bias and by modifying the impact parameter.
These analyses used the SIS lens model and thus neglected
inclination effects in the lensing properties, although
they did include inclination effects in the HI column
density.  The strong inclination dependence of the lensing
properties must be taken into account in order to properly
treat the effects of lensing on the statistics of damped
Ly$\alpha$ absorbers.

Finally, our results suggest that spiral galaxies cannot
explain the weak discrepancy between observed lens galaxy
axis ratios and the model axis ratios required to explain
individual lenses and the statistics of 4-image lenses
(King et al.\ 1996; Kochanek 1996b; Keeton et al.\ 1997a).
While it is true that edge-on spiral galaxies can produce
many 4-image lenses, the absence of observed ``disk'' lenses
indicates that spirals do not contribute significantly to
the observational sample.  Moreover, the disk does not
substantially increase the total cross section compared
with the SIS model.  Thus the only way to increase the
fraction of lenses due to spirals is to change the ratio
of spiral to early-type galaxy number densities.  Kauffmann,
Charlot \& White (1996) have offered evidence that evolution
may reduce the number of early-type galaxies by as much as
a factor of 2--3 at $z=1$.  However most lens galaxies are
closer than $z=1$, and it seems unlikely that number
evolution could change the ratio of spirals to early-types
by the factor of $\sim5$--$10$ that would be required to
make spiral galaxies dominate lens samples.

\acknowledgments
Acknowledgements:
We thank A.~Loeb for useful discussions.
CRK is supported by ONR-NDSEG grant N00014-93-I-0774.
CSK is supported by NSF grant AST-9401722 and NASA ATP
grant NAG5-4062.


\clearpage


\begin{deluxetable}{rcccc}
\tablecaption{Scale Length Ratio $a_h/a_d$ for a Flat Rotation Curve}
\tablehead{  & $\qq{d}=0.1$ & $0.03$ & $0.01$   & $0.0$ }
\startdata
$\qq{h} = 1/3$ & $0.9091$ & $0.8801$ & $0.8717$ & $0.8675$ \\
			$1/2$ & $0.8569$ & $0.8296$ & $0.8217$ & $0.8177$ \\
				$1$ & $0.7397$ & $0.7161$ & $0.7092$ & $0.7058$ \\
\enddata
\tablecomments{Values of the ratio $a_h/a_d$ needed to produce a
flat rotation curve in a model with a truncated Mestel disk in an
softened isothermal halo, where $\qq{d}$ and $\qq{h}$ are the axis
ratios of the disk and halo, respectively.  These values give a
rotation curve that is flat to better than $2\%$.}
\label{tab:magic}
\end{deluxetable}

\clearpage


\end{document}